\documentclass[]{article}
\usepackage{cite}
\usepackage{amsmath,amssymb,amsfonts}
\usepackage[noend]{algpseudocode}
\usepackage{algorithmicx,algorithm}
\usepackage{graphicx}
\usepackage{textcomp}
\usepackage{xcolor}
\usepackage{booktabs}
\usepackage{multirow}
\usepackage{float}
\usepackage{bigstrut}

\newtheorem{definition}{Definition}

\def\BibTeX{{\rm B\kern-.05em{\sc i\kern-.025em b}\kern-.08em
		T\kern-.1667em\lower.7ex\hbox{E}\kern-.125emX}}
\date{}

\begin{document}
\title{ECG Classification based on Wasserstein Scalar Curvature
 }
\author{
Fupeng Sun,~ Yin Ni,~ Yihao Luo,~ Huafei Sun\thanks{This research was funded by National Key Research and Development Plan of China, No.
		2020YFC2006201}\\
}
\maketitle{}
\large

\noindent\textbf{Abstract}: Electrocardiograms (ECG) analysis is one of the most important ways to diagnose heart disease. This paper proposes an efficient ECG classification method based on Wasserstein scalar curvature to comprehend the connection between heart disease and mathematical characteristics of ECG. The newly proposed method converts an ECG into a point cloud on the family of Gaussian distribution, where the pathological characteristics of ECG will be extracted by the Wasserstein geometric structure of the statistical manifold. Technically, this paper defines the histogram dispersion of Wasserstein scalar curvature, which can accurately describe the divergence between different heart diseases. By combining medical experience with mathematical ideas from geometry and data science, this paper provides a feasible algorithm for the new method, and the theoretical analysis of the algorithm is carried out. Digital experiments on the classical database with large samples show the new algorithm's accuracy and efficiency when dealing with the classification of heart disease.

\noindent\textbf{keywords}: ECG Classification, Positive definite symmetric matrix manifold, Wasserstein metric, Curvature

\section{Introduction}
As one of the most common and deadly diseases in the world, heart disease poses a significant threat to people's happiness and healthy life. With the increase of social pressure, the incidence of heart disease is proliferating in recent years\cite{WHO}. Therefore, it is of great indispensability to realize the efficient diagnosis, real-time monitoring and prediction for heart disease. At present, the diagnosis is mainly made through doctors' manual analysis of ECG (electrical activity records of patient's heart contraction)\cite{ECG}. With no exception, ECG analysis requires doctors to command professional and detailed medical knowledge. However, as a kind of valuable medical resource, the distribution of experienced doctors is not well balanced in different regions of the world. Thus, the life and health of patients can not be fully protected in the situation with poor medical conditions and resources.

With the rapid development of computer technology, using computer-aided diagnosis technology to screen heart diseases has become a new technical method to alleviate the imbalance of artificial medical resources. The normal ECG shows regular changes of PQRST complex \cite{PQRST}, and different pathological features will cause symmetry vanishing. The computer-aided diagnosis techniques aim to extract different features of ECG and detect fluctuation of PQRST complex in the view of these features to achieve accurate classification. 

Commonly used computer-aided diagnosis techniques mainly extract features from the viewpoints of signal analysis, dynamic system modeling (DSA) and topological data analysis (TDA), which are also combined with classic statistical analysis\cite{Stamkopoulos T} and machine learning\cite{Li Jianning,Weijia Lu,Yan Y}. Signal analysis can directly extract morphological features such as amplitude\cite{Christov I,Ye C} or use wavelet transform to acquire frequency domain features\cite{Banerjee S,He R}. Other emerging signal analysis methods\cite{Hua X} also have the potential to analyze ECG. Dynamical system analysis is used to model cardiac dynamical system through phase space reconstruction\cite{Richter M,Chen C K,Desai U,Di Marco L Y,Gong Y}. Topological data analysis transforms ECG signals into point clouds by time-delay embedding or Fourier transform, and then extracts topological features from point clouds\cite{Fraser B A,Safarbali B,Ignacio P S,Graff G,Dindin M,Ni Yin}. Above mentioned methods are adaptive with different applied conditions. Despite various advantages of conventional methods, most of them are limited by small sample size or poor interpretation. Thus a new way without such limitations need to be proposed.

Inspired by persistent homology (PH)\cite{U Bauer} in TDA, a geometric data analysis method is proposed and applied to ECG classification. PH transforms the original ECG signals into point clouds in Euclidean space by embedding in a suitable way and extracts topological features from point clouds. Topological features can reflect different pathological
characteristics of the original ECG, so as to realize the classification. But the topological characteristics cannot focus on fine local information of point clouds, also the category number and sample size are limited. Thus, it is necessary to describe the local differences of point clouds in a more accurate method.

Each point of the point cloud in $n$ dimensional Euclidean obtained by Fast Fourier transform embedding (FFT embedding) \cite{Ni Yin} is mapped to a $n$ dimensional Gaussian distribution by local statistic, that is, the point cloud in Euclidean space is transformed into the point cloud on the positive definite symmetric matrix manifold $SPD(n)$. $SPD(n)$ becomes a Riemannian manifold after being endowed with the natural Riemannian metric induced by Wasserstein distance \cite{Rubner Y}. We reveal the local differences of point clouds on $SPD(n)$\cite{Luo Y} by Wasserstein scalar curvature (WSC). By analyzing Wasserstein scalar curvature histogram (WSCH) of the converted point clouds, we originally defined a new data characteristic, called Wasserstein scalar curvature dispersion (WSCD) which can express the pathological changes of heart. Finally, the ECG classification algorithm based on Wasserstein scalar curvature (WSCEC) is designed to classify ECG signals. 

This paper is organized as follows. Section \ref{section 2} gives the preliminaries. Section \ref{section 3} presents WSCEC algorithm. In Section \ref{section 4}, we analyze the numerical results. Finally, we discuss the main conclusions and prospects for the future research in Section \ref{section 5}.

\section{Preliminaries}\label{section 2}
In this section, we will introduce some basic preliminaries such as FFT embedding, local statistic and Wasserstein geometric structure on $SPD(n)$.

\subsection{Fast Fourier transform embedding}
Fourier transform decomposes signals into waves with different frequencies and reveals the certain features hidden in the time domain. For discrete inputs, fast Fourier transform (FFT) is a widely used tool. Given a signal $T=\{t_i\}_{i=1}^n$ with even $n$, $t_i$ can be represented as:
\begin{equation}\label{fft1}
	t_i=\frac{1}{n}\sum_{k=0}^{n-1}C_k e^{j\frac{2\pi}{n}ki},
\end{equation}
where $C_k=\sum_{i=0}^{n-1}t_ie^{-j\frac{2\pi}{n}ki}$.

Let $C_k=a_k+b_kj$, then we have $C_{n-k}=a_k-b_kj$ and
\begin{equation}\label{fft2}
	C_{k}e^{j\frac{2\pi}{n}ki}+C_{n-k}e^{j\frac{2\pi}{n}(N-k)i}=A_kcos\left(\frac{2\pi ki}{n}+\phi_k\right),
\end{equation}
where $A_k=2\sqrt{a_k^2+b_k^2}$, $tan(\phi_k)=\frac{b_k}{a_k}$.

If the sample frequency is $f_s$, by combining equation (\ref{fft1}) and equation (\ref{fft2}), we have
\begin{equation}
	t(s)=\frac{1}{n}\sum_{k=0}^{\frac{n}{2}-1}A_kcos\left(f_s\frac{2\pi ks}{n}+\phi_k\right),
\end{equation}
where $t_i=t\left(\frac{i}{f_s}\right)$.

Now we use FFT and sliding windows to convert signal $T=\{t_i\}_{i=1}^n$ to a point cloud in $\mathbb{R}^{d}$. Let $l$ be window length and $\tau$ be sliding speed. Firstly, we transform $T$ into a siganl set $\mathcal{P}_{l,\tau}(T)=\{p_k\}_{k=1}^{\hat{n}-1}$ with $\hat{n}=\left[ \frac{n-l}{\tau}\right]$ and $p_k=\left[t_{k\tau},t_{k\tau+1},\cdots,t_{k\tau+(l-1)}\right]$. Secondly, we transform each $p_k$ into a point in $\mathbb{R}^{d}$ by choosing the first $d$ bases from FFT and obtain the following point cloud:
\begin{equation}\label{fftpoint}
	T\rightarrow \mathcal{P}_{l,\tau}(T)=\{p_k\}_{k=1}^{\hat{n}-1}\rightarrow S_{F}(T)=S_{l,\tau,d}(T)=\left\{\frac{2}{l}\left(a_0^k,a_1^k,b_1^k,\cdots,a_{\frac{d}{2}-1}^k,b_{\frac{d}{2}-1}^k\right)\right\}_{k=1}^{\hat{n}-1}.
\end{equation}  
where $d\leq l$.

\subsection{Local statistic}
Objective phenomena in nature are often disturbed by many small random variables, which assigns a random distribution with the local neighborhood of any point in the point cloud. If the factors which affect the local distribution of a point cloud are considered small and complex enough, then according to the law of large numbers and the central limit theorem, we can assume that the local statistic comes from a high-dimensional Gaussian distribution whose parameters are neighborhood mean and neighborhood covariance matrix.

In data science, $kNN$ algorithm provides a natural neighborhood selection method. The idea is to search for the nearest k points as the neighborhood samples of any fixed point in the point cloud. To acquire local statistic, for every point $P_i\in S_{F}(T)$, we can search $kNN$ to obtain neighborhood $\mathcal{N}_{i}=\{N_{ij}|j=1,\cdots,k\}$, and calculate local mean $\mu_i$ and covariance matrix $\Sigma_i$:
\begin{equation}\label{local mean}
	\mu_i = \frac{1}{k}\sum_{j=1}^{k}N_{ij},~\Sigma_i = \sum_{j=1}^{k}(N_{ij}-\mu_i)^T(N_{ij}-\mu_i).
\end{equation}
Consequently, $S_{F}(T)$ is converted to a point cloud in $SPD(d)$:
\begin{equation}\label{local statistic}
	S_{F}(T)\rightarrow S_{P}(T)=S_{SPD}(T,k)=\left\{ \Sigma_i \mid P_i\in S_{F}(T)  \right\}.
\end{equation}

\subsection{Wasserstein geometric structure on $SPD(n)$}
Wasserstein distance describes the minimal energy used to transport one distribution to another. It can be used to measure the difference between two distributions and is vividly called earth-moving distance\cite{Rubner Y}.

Let $F_1,F_2$ be two distributions. Then the Wasserstein distance between $F_1$ and $F_2$ is defined as the infimum of geodesic distance integral needed for transporting probability measure element:
	\begin{equation}\label{Wsst}
		W_p\left(F_1,F_2\right)= \inf_{p\sim \Pi\left(F_1,F_2\right)} \left(E_{\left(x,y\right)\sim p}[\|x-y\|^p] \right)^{\frac{1}{p}},
	\end{equation}
where $\Pi\left(F_1,F_2\right)$ is the set of joint distributions taking $F_1,F_2$ as marginal distributions, $E$ is the expectation.

Although the definition is abstract and there is no explicit expression for the general Wasserstein distance, the Wasserstein distance between any two Gaussian distributions in $\mathbb{R}^n$ has the following explicit expression\cite{Givens C}:
\begin{equation}\label{Wss2d} 
   W\left(\mathcal{N}_1,\mathcal{N}_2\right)=\|\mu_1-\mu_2\|+ \left({\rm tr}\left(\Sigma_1+\Sigma_2-2\left(\Sigma_1\Sigma_2\right)^{\frac{1}{2}}\right)\right)^{\frac12},
\end{equation}
where $\mu_i$ and $\Sigma_i$ are the mean and covariance matrix of Gaussian distribution $\mathcal{N}_i$, $i=1,2$.

Wasserstein distance can be induced by a Riemannian metric on $SPD(n)$ defined as:
\begin{equation}\label{Gw metric}
	g_W|_{S}\left(X,Y\right) = \frac{1}{2}{\rm tr}\left(\Gamma_{S}[Y]  X\right),
\end{equation}
where ${S}\in SPD\left(n\right)$, $X,Y\in T_{S}SPD\left(n\right)$ are tangent vectors and $\Gamma_{S}[Y]$ is the solution of Sylvester equation $S\Gamma_{S}[Y]+\Gamma_{S}[Y]S=Y$\cite{Ward A}.

Because the geodesic distance induced by the equation \eqref{Gw metric} is consistent with the original definition of Wasserstein distance \eqref{Wss2d}, we call $g_W$ Wasserstein metric. In addition, we write the Riemannian manifold $SPD(n)$ endowed with Wasserstein metric as $(SPD(n),g_W)$.

For any ${S} \in (SPD(n),g_W)$, let ${X},{Y}$ be the smooth vector field of $SPD\left(n\right)$. \cite{Massart E, Luo Y} provide the explicit expression of Riemannian curvature tensor $\langle R_{{X}{Y}}{X},{Y}\rangle$ at ${S}$:
\begin{equation}\label{Riecurvature}
	\begin{split}
		R\left({X},{Y},{X},{Y}\right)
		=3{\rm tr}  & \left(\Gamma_{S}[{X}]{S}\Gamma_{S}\left(\Gamma_{S}[{X}]\Gamma_{S}[{Y}]-\Gamma_{S}[{Y}]\Gamma_{S}[{X}]\right){S}\Gamma_{S}[{Y}]\right).
	\end{split}
\end{equation}
Furthemore, the scalar curvature at ${S}$ satisfies
\begin{equation}\label{Curve scalar}
	\begin{split}
		\rho\left({S}\right) &= \sum_{i=1}^{n}\sum_{j=1}^{n}R\left({e_i},{e_j},{e_i},{e_j}\right)\\
		&= 3{\rm tr}\left( U\Lambda (U + U^{T})+ (U + U^{T})\Lambda U+ (U + U^{T})\Lambda U\Lambda \left(U+U^{T}\right)\right),
	\end{split}
\end{equation}
where $\{e_i\}$ is any standard orthonormal basis of $T_S{SPD(n)}$, $\Lambda= {\rm diag}\left(\lambda_{1}, \cdot\cdot\cdot, \lambda_{n}\right)$ is orthogonal similar to ${S}$, $\lambda_{i}$ is the $i$th eigenvalue of ${S}$, and $U=\left[\frac{1}{\lambda_{i}+\lambda_{j}}\right]_{i<j}$ is an upper triangular matrix.

Note that the scalar curvature of ${S}$ can be controlled by the second small eigenvalue of ${S}$. Actually, there exists a standard orthonormal basis $\{e_k\}$ of $T_{S}SPD\left(n\right)$, such that $\forall$ $e_{k_1},e_{k_2} \in \{e_k\}$,
	\begin{equation} 0<K_{S}\left(e_{k_1},e_{k_2}\right)=\sum_{j=1}^{n}R\left({e_{k_1}},{e_{k_j}},{e_{k_2}},{e_{k_j}}\right)< \frac{3}{\lambda_{min2}\left({S}\right)},
	\end{equation}
where $\lambda_{min2}$ is the second small eigenvalue of ${S}$.
	
Furthemore, by equation (\ref{Curve scalar}), we have
	\begin{equation}\label{cur inequality}
		0<\rho\left({A}\right)< \frac{3n(n-1)}{\lambda_{min2}\left({S}\right)}.
	\end{equation}

The boundedness of WSC indicates that the curvature of local covariance matrix is controllable unless the local covariance degenerates in two dimensions or more. Consequently, for most neighborhoods, as long as using appropriate embedding methods to make sure there is no degeneration beyond two dimensions, WSC of point cloud on $SPD(n)$ will be in a controllable range, which provides a theoretical criterion for the robustness of our algorithm.

\section{ECG Classification Algorithm based on Wasserstein Scalar Curvature}\label{section 3}
\begin{figure}[htbp]
	\centerline{\includegraphics[width=1\linewidth]{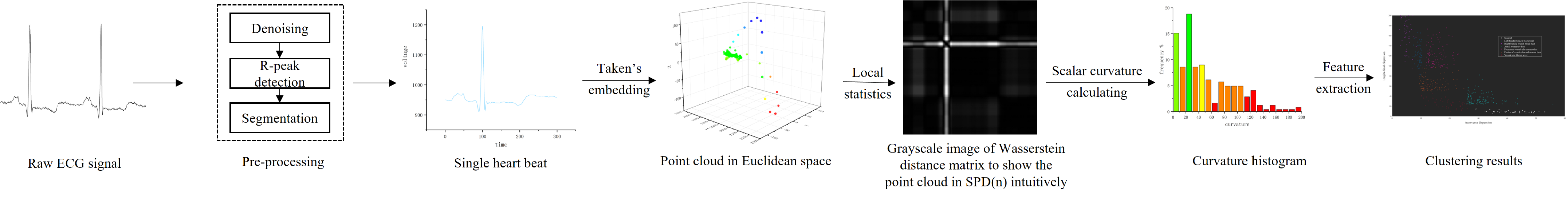}}
	\caption{Pipeline of WSCEC}
	\label{pipeline}
\end{figure}

In this section, we will introduce WSCEC algorithm which can detect the heart disease. Wasserstein scalar curvature dispersion is extracted to reveal the change of regularity of PQRST complex. The framework of WSCEC algorithm is as follows.
\begin{itemize}
	\item Continuous ECG signals are segmented and denoised by interpolation and filter to obtain multiple single heartbeats.
	\item Every single ECG is transformed into a point cloud in Euclidean space by FFT embedding. Through local statistic, the point cloud in Euclidean space is converted to the point cloud on $SPD(n)$.
	\item Calculate WSC of each point to obtain WSCH and extract WSCD as the feature.
	\item Do auxiliary diagnosis according to clustering results.
\end{itemize}
The intuitive algorithm pipeline is shown in Figure \ref{pipeline}.

\subsection{Preprocessing of ECG signal}
We adopt the idea of Butterworth filter algorithm\cite{Butterworth S} to cut off noisy portions with spectral power over 50 Hz. By developing a local search algorithm to find the periodic R-peak in PQRST complex, we successfully transform continuous ECG signals into multiple single heartbeats with length $300$.

\begin{figure}[htbp]
	\setlength{\abovecaptionskip}{-0.5cm}
	\centerline{\includegraphics[width=0.8\linewidth]{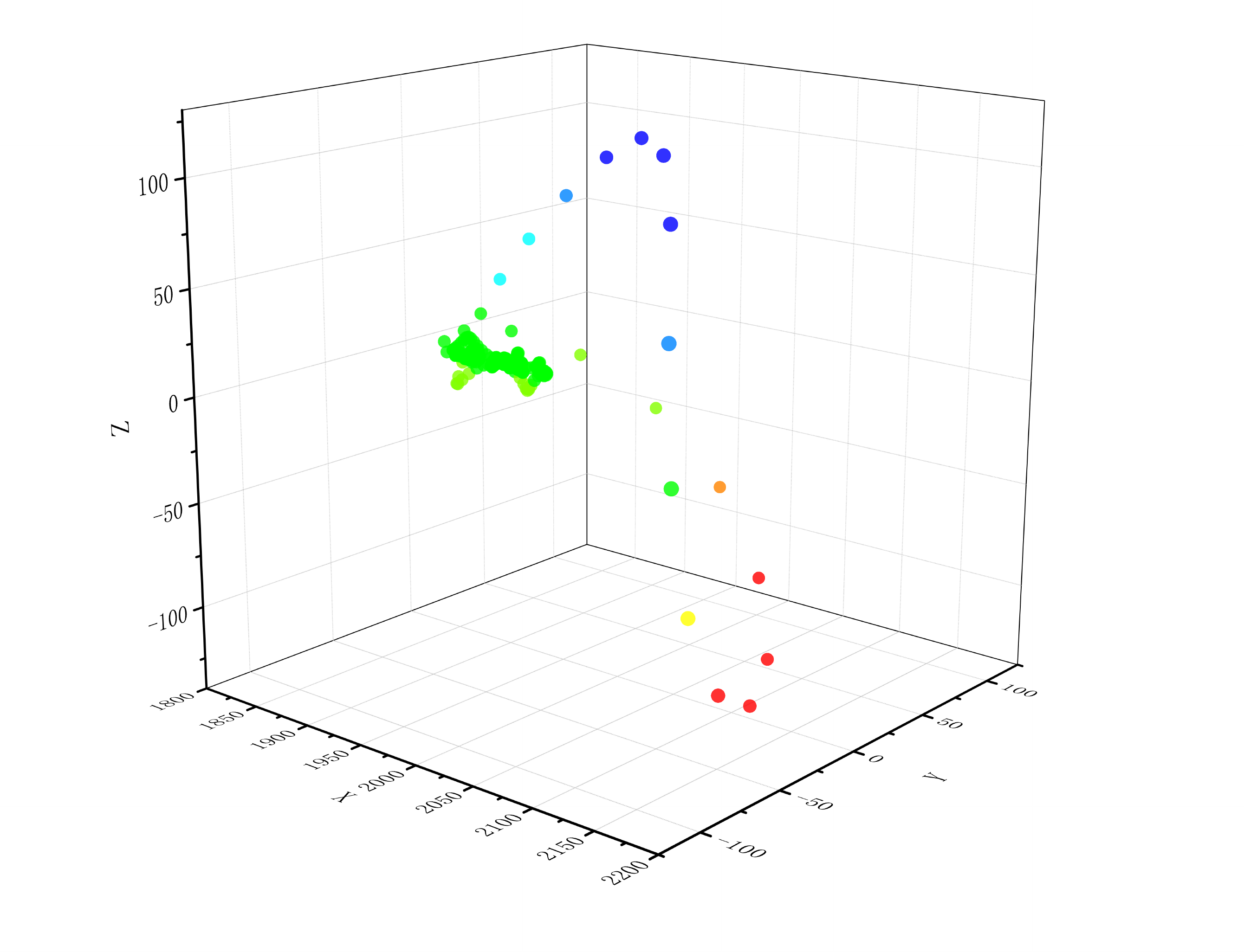}}
	\caption{Point cloud of $T_{s}$ in $\mathbb{R}^3$ after FFT embedding}
	\label{takens1}
\end{figure}

Noting that the length of the sharp part of QRS complex in normal ECG is around $10$, we set window length $l=10$ and sliding speed $\tau=1$ to emphasize the change of QRS complex when sliding the window. Then by using equation (\ref{fftpoint}) and taking $d=3$, we convert various single heartbeats to point clouds in $\mathbb{R}^3$. Let $T_ {s}$ be a standard normal ECG signal, denote $S_{F}(T_{s})$ as the Euclidean point cloud of $T_ {s}$ after FFT embedding. Figure \ref {takens1} shows $S_{F}(T_{s})$ intuitively.

\begin{figure}[htbp]
	\centerline{\includegraphics[width=0.6\linewidth]{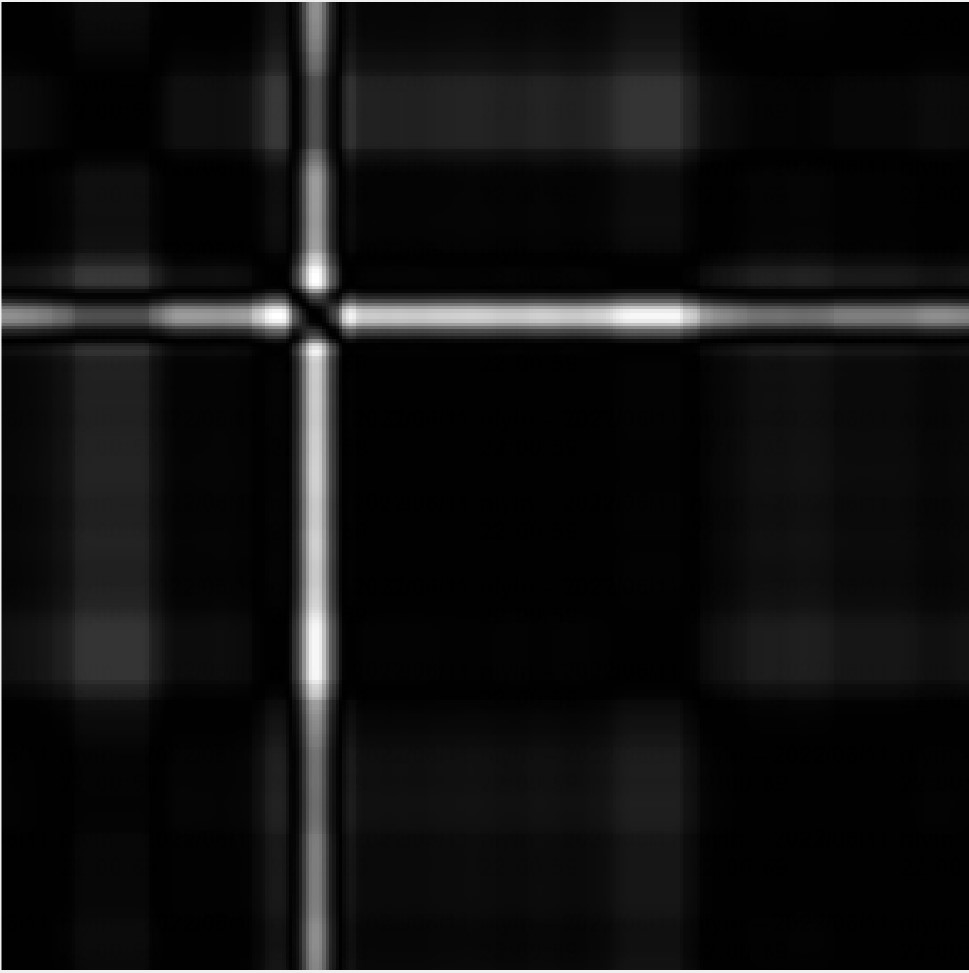}}
	\caption{Grayscale image of Wasserstein distance matrix for point cloud of $T_{s}$ on $SPD(3)$}
	\label{wassfigure1}
\end{figure}

In attempt to describe the local differences of Euclidean point clouds more accurately, we obtain neighborhood properties by local statistic. We combine $kNN$ algorithm with equation (\ref{local mean}) to obtain $S_{P}(T_{s})$, the parameter of $kNN$ algorithm is $k=20$. With an attempt to give readers an intuitive understanding of the structure of point clouds on $SPD(3)$, we acquire Wasserstein distance matrices by equation (\ref{Wss2d}) and present the grayscale images of Wasserstein distance matrice for $T_{s}$, see Figure \ref{wassfigure1}.

\subsection{Feature extraction}
The pathological differences of original ECG signals are completely reflected by the local information of point clouds, and the differences of local information are further reflected in the neighborhood mean and covariance matrix, that is, reflected in the point-by-point difference of point clouds on $SPD(n)$. Since Wasserstein scalar curvature reflects the structural relationship between a point and its adjacent points, we can use WSC to describe different pathological features of ECG signals.

Firstly, we calculate WSC of each point in the distribution point cloud, then the corresponding WSC sequence $\mathcal{W}=\{w_i\}_{i=1}^{\hat{n}}$ can be obtained. Furthemore, we can acquire WSCH:
\begin{equation}\label{histogram}
	H(m,b)=\left\{\left([mj,m(j+1)),y_j\right)\mid 0\leq j \leq \left[\frac{b}{m} \right]\right\},
\end{equation}
where
\begin{equation}
	y_j=\mid \left\{w_k\in \mathcal{W}\mid mk\leq w_k < m(k+1) \right\}\mid,
\end{equation}
	$\left[\cdot\right]$ denotes integer operator and $|\cdot|$ denotes the cardinality or size of a finite set.

\begin{figure}[htbp]
	\setlength{\abovecaptionskip}{-0.3cm}
	\centerline{\includegraphics[width=0.9\linewidth]{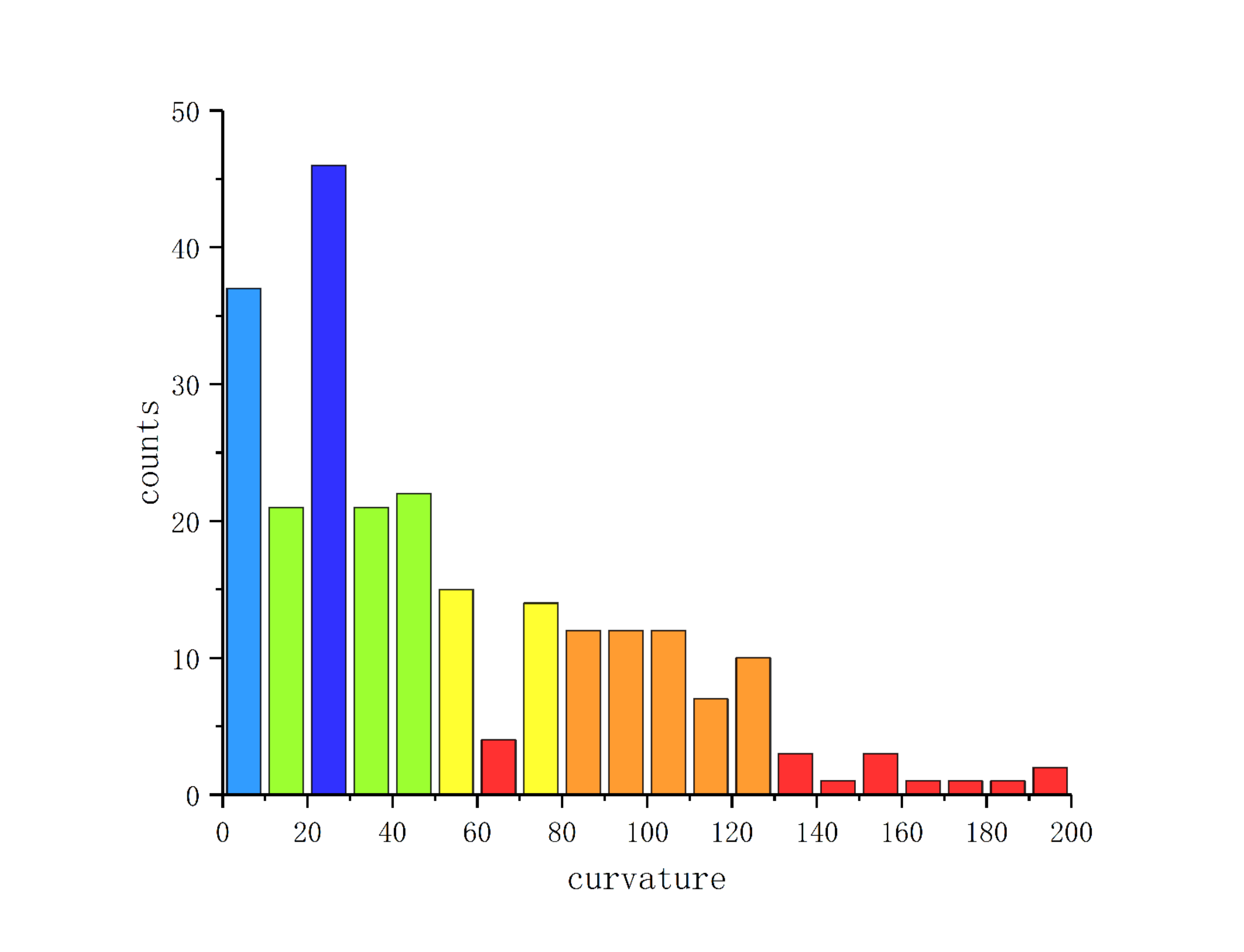}}
	\caption{WSCH for distribution point cloud of $T_{STD}$}
	\label{fig:histogram1}
\end{figure}

Figure \ref{fig:histogram1} performs the WSCH of the point cloud $S_{P}(T_{s})$ on $SPD(3)$ where $m=10, b=200$.
Now we give the definition of WSCD to describe the differences of WSC sequence.

\begin{definition}
	For the given WSC sequence $\mathcal{W}$, its histogram $H(m,b)$ and $0\leq s \leq \left [\frac{b}{m}\right]~(s\in \mathbb{N}_{+})$, define:
	\begin{equation}\label{cureq}
			\begin{aligned}
					cur_1(m,b,s)&= median \ of \ \mathcal{U}_1, \\
					cur_2(m,b,s)&=\frac{1}{\mid \mathcal{U}_2 \mid-s-1}\sum_{j\geq s+1} \left( y_j-\frac{\sum_{j\in \mathcal{U}_2}y_j}{\left[\frac{b}{m} \right]}\right)^2, \\
					cur(m,b,s)&=(cur_1(m,b,s),cur_2(m,b,s)),
				\end{aligned}
		\end{equation}
where
\begin{equation}
	\begin{aligned}
				\mathcal{U}_1&=\left\{w_k\in \mathcal{W}\mid ms\leq w_k\leq b\right\},\\
			\mathcal{U}_2&={\left\{k\geq s+1\mid y_k\neq0 \right\}}.
		\end{aligned}
\end{equation}
We call $cur_1(m,b,s)$ Wasserstein scalar curvature transverse dispersion of $\mathcal{W}$, $cur_2(m,b,s)$ Wasserstein scalar curvature longitudinal dispersion of $\mathcal{W}$ and cur(m,b,s) Wasserstein scalar curvature dispersion of $\mathcal{W}$, respectively. 
\end{definition}
In histogram $H(m,b)$, transverse dispersion $cur_1(m,b,s)$ is the median of the intersection between $\mathcal{W}$ and $[ms,b]$. $cur_1(m,b,s)$ describes the homogeneity of $\mathcal{W}$ ranging in $[m(s+1),b]$ horizontally. 

The longitudinal dispersion $cur_2(m,b,s)$ represents the fluctuation of the column, which can be regarded as the correction of the standard deviation. If the column heights of some curvature intervals in $H(m,b)$ are $0$, then this fluctuation is further amplified. $cur_2(m,b,s)$ describes the uniformity of $\mathcal{W}$ ranging in $[m(s+1),b]$ vertically. In particular, $cur_2(m,b,0)$ is the standard deviation of the curvature histogram $H(m,b)$ if the columns are all nonzero.

\begin{figure}[htbp]
	\centerline{\includegraphics[width=1\linewidth]{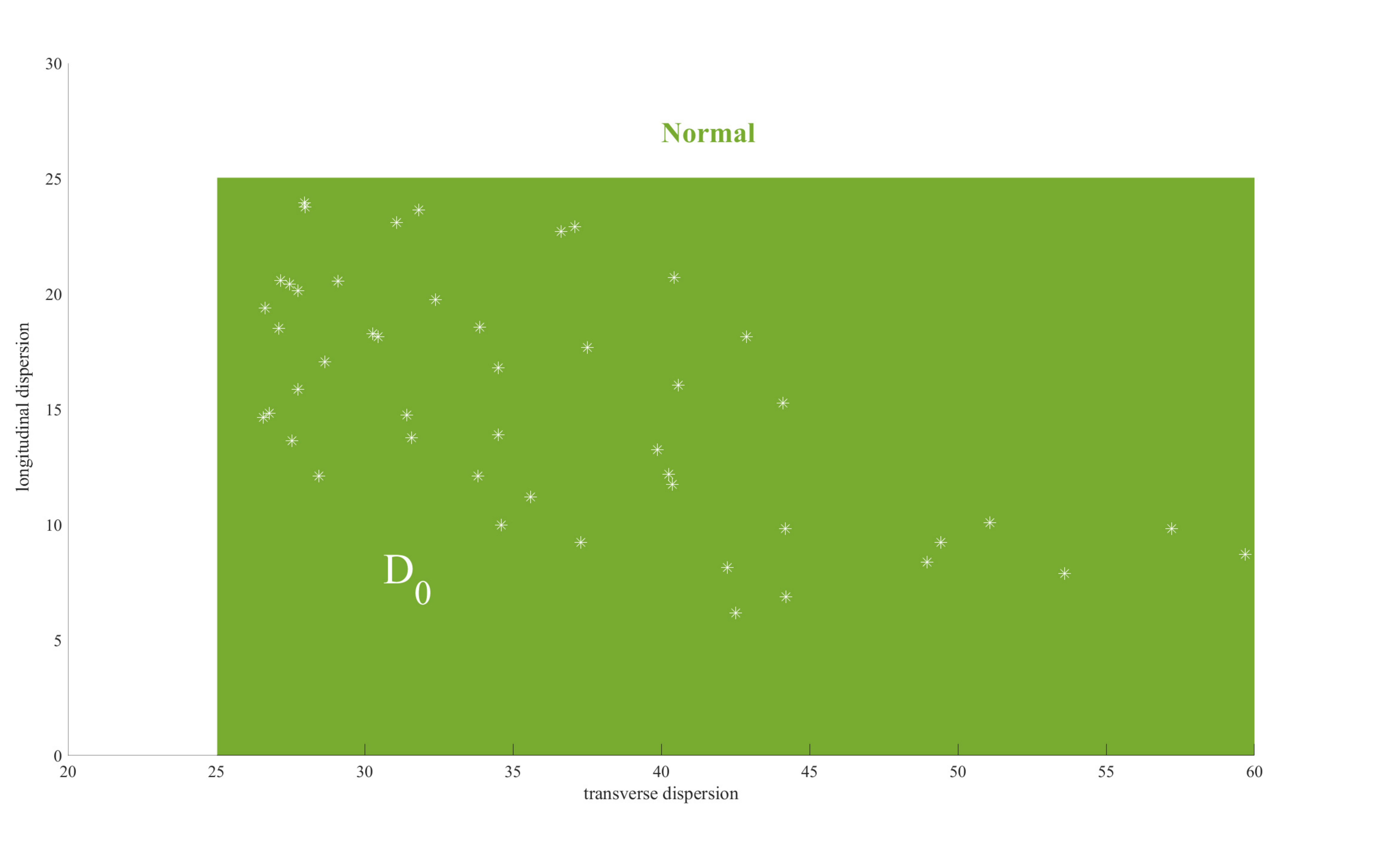}}
	\caption{WSCD of normal ECG signals}
	\label{fig:normalfeature}
\end{figure}

Therefore, $cur(m,b,s)$ describes the homogeneity of $\mathcal{W}$ overall, $cur(m,b,s)$ is closer to $\left(\frac{1}{2}(b+ms),0\right)$ if the columns are more evenly distributed. Since the ECG of healthy heartbeats have strong regularity, their WSCD are more even, as shown in Figure \ref{fig:normalfeature}. We consider WSCD $cur(m, b, s)$ as the feature of our final classification.

\subsection{Case analysis}
\begin{figure}[htbp]
	\vspace{-0.3cm}
	\setlength{\abovecaptionskip}{-0.8cm}
	\centerline{\includegraphics[width=0.7\linewidth]{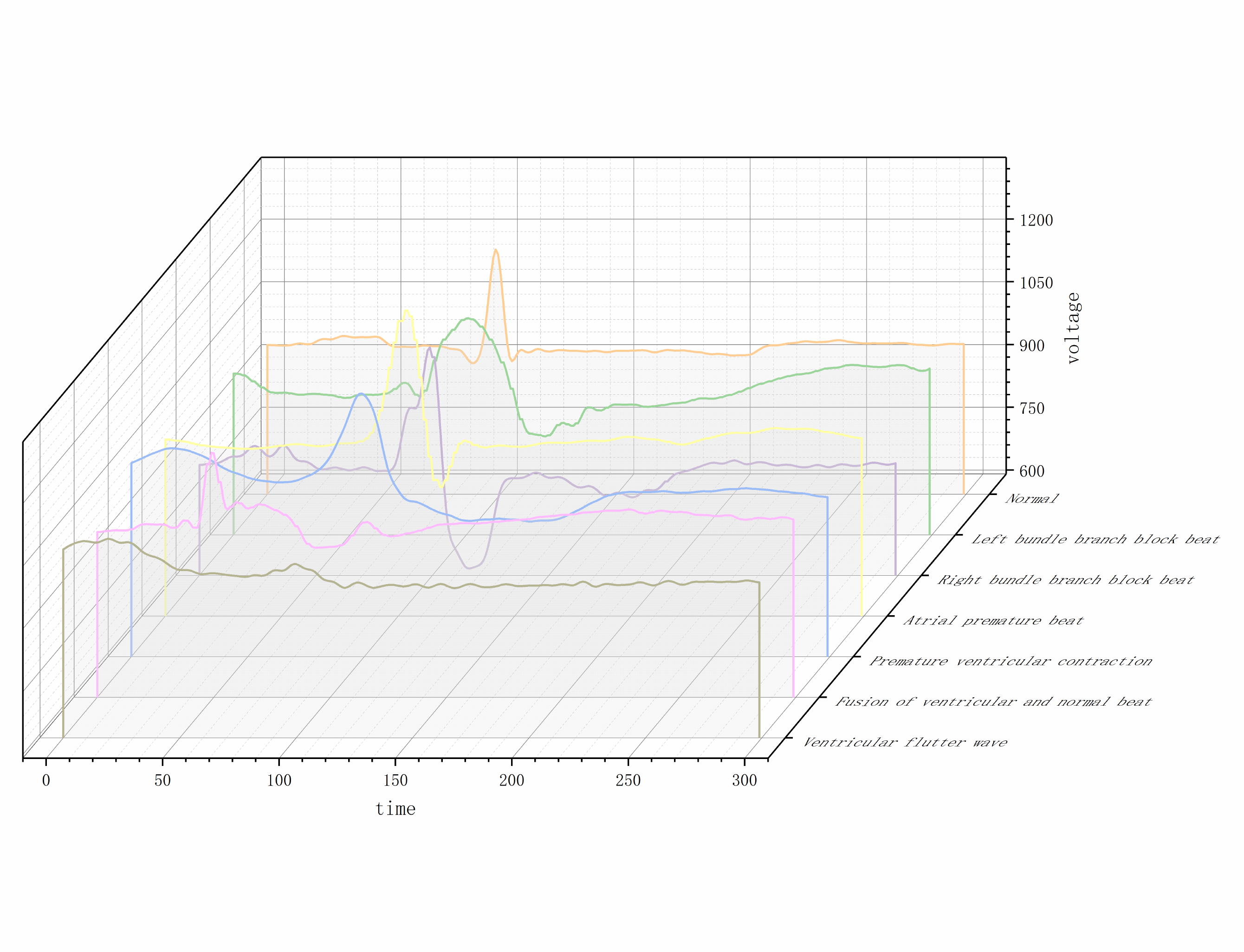}}
	\caption{Schematic diagram of seven kinds of ECG signals}
	\label{fig:ecg}
\end{figure}

Figure \ref{fig:ecg} shows seven types of single heartbeats. Compared with normal ECG signal, other six diseases have different effects on PQRST complex. The QRS complex of left bundle branch block heartbeats (L. B. B. B.) or right bundle branch block heartbeats (R. B. B. B.) is obviously broadened, and generally there are two R peaks. The P wave of atrial premature heartbeats (A. P.) occurrs earlier and is significantly different from that of sinus. P. V. C. has the larger QRS complex amplitude, which always companies with more significant range differences in the waves. The QRS complex in fusion of ventricular and normal heartbeats (F. V. N.) is the fusion of normal heartbeat and ventricular flutter heartbeats (V. F.), and its deterioration will change into V. F. whose waveform is similar to sine wave, in which case cardiopulmonary resuscitation is needed for treatment.

\begin{figure}[htbp]
	\centerline{\includegraphics[width=1\linewidth]{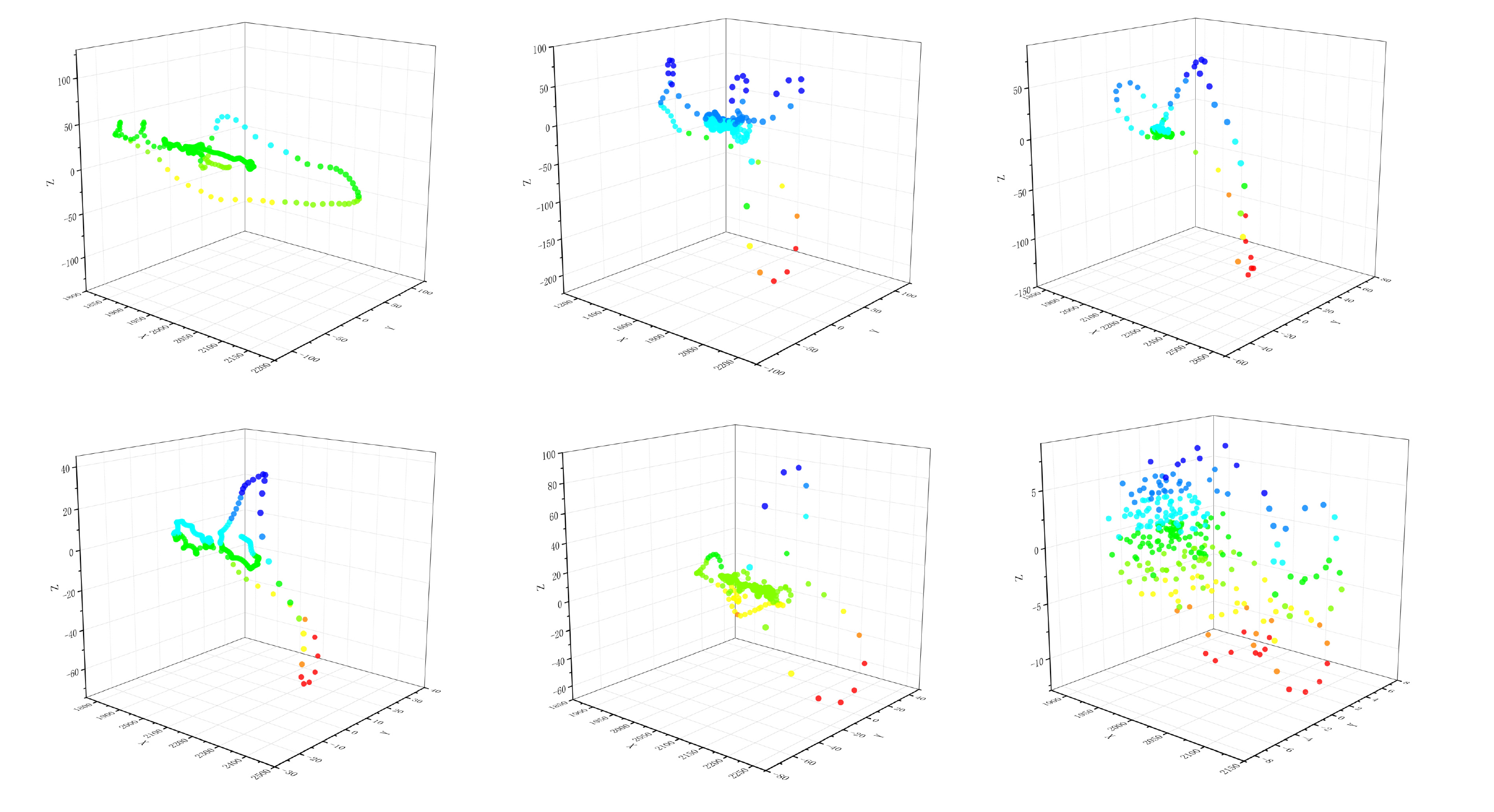}}
	\caption{Point clouds of ECG signals with six pathological features in $\mathbb{R}^3$ after FFT embedding; the first row from left to right represents L. B. B. B., R. B. B. B., A. P.; the second row from left to right represents P. V. C., F. V. N., V. F.}
	\label{fig:ecgothers}
\end{figure}

Figure \ref{fig:ecgothers} shows the point clouds in $\mathbb{R}^3$ for ECG signals with six pathological features. Compare Figure \ref{takens1} and Figure \ref{fig:ecgothers}, the differences of waves are reflected in the differences of local information of point clouds in $\mathbb{R}^3$. Notice that the local information of the point clouds of ECG signals with diseases are significantly different from those of normal ECG signals except A. P., this may be due to the fact that A. P. is generated by atrial abnormal excitation foci in advance, and sometimes there are only $P$ wave differences with normal ECG signals. Therefore, their local structures of Euclidean point clouds are similar. 

There are also similarities among the point cloud structures of P. V. C., F. V. N., and V. F., which may be due to the fact that these three types of diseases are also generated from ventricular ectopic excitation foci, their pathogenesis and trend also have a certain progressive relationship. The similarities of local structures between different ECG signal point clouds also reflect the necessity of introducing WSC to describe the differences of such fine structures more accurately.

\begin{figure}[htbp]
	\centerline{\includegraphics[width=0.9\linewidth]{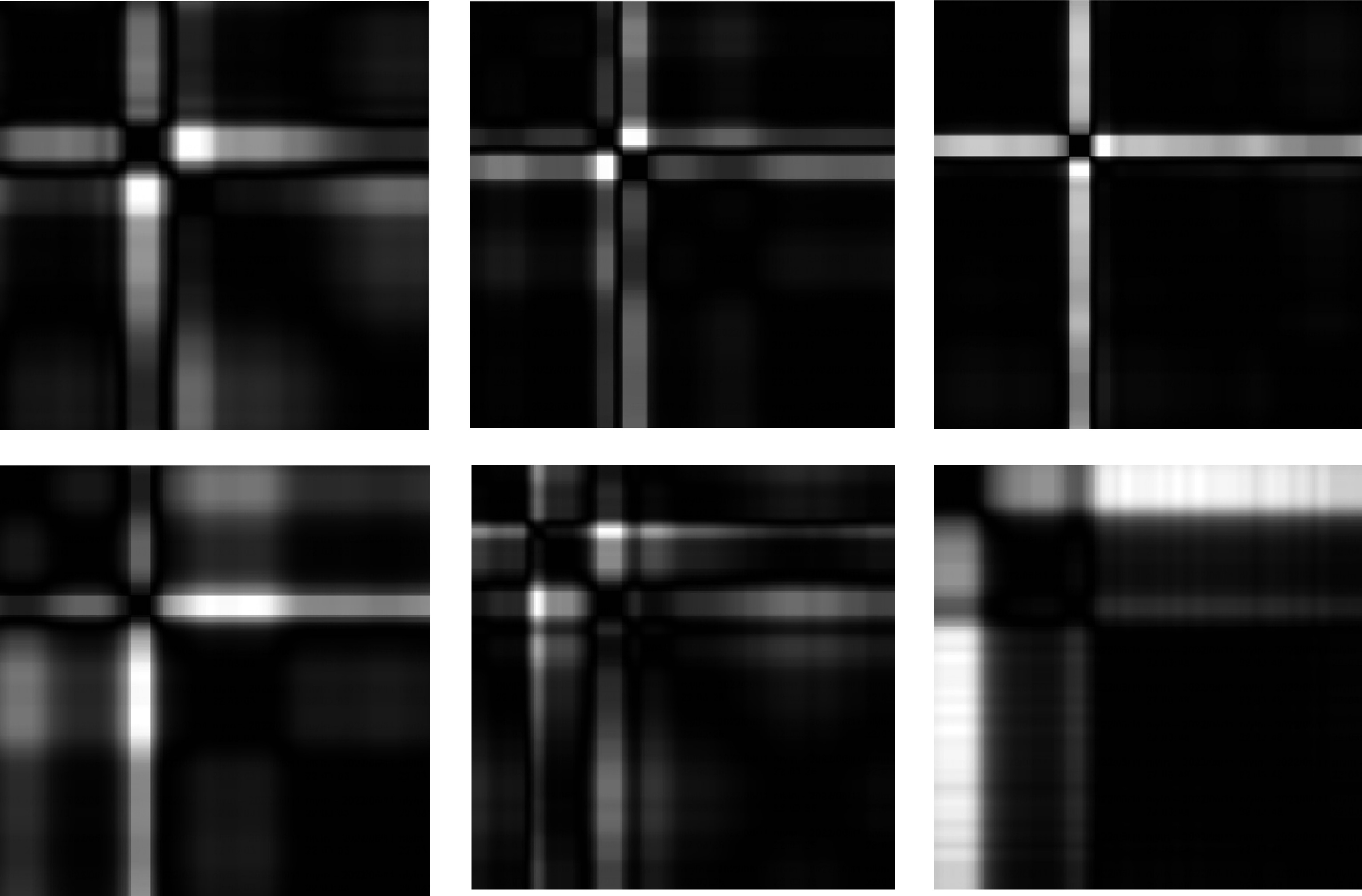}}
	\caption{Grayscale image of Wasserstein distance matrices for distribution point clouds; the first row from left to right represents L. B. B. B., R. B. B. B., A. P.; the second row from left to right represents P. V. C., F. V. N., V. F.}
	\label{fig:dmatrixothers}
\end{figure}

By local statistic, we change the point clouds of ECG signals with diseases into the point clouds on $SPD(3)$. The grayscale images of Wasserstein distance matrices describe the dispersion of distribution point clouds on $SPD(3)$, where black represents the zero distance and white represents the maximum distance. Local structure differences of distribution point clouds can be visually presented by Figure \ref{wassfigure1} and Figure \ref{fig:dmatrixothers}.

\begin{figure}[htbp]
	\centerline{\includegraphics[width=1\linewidth]{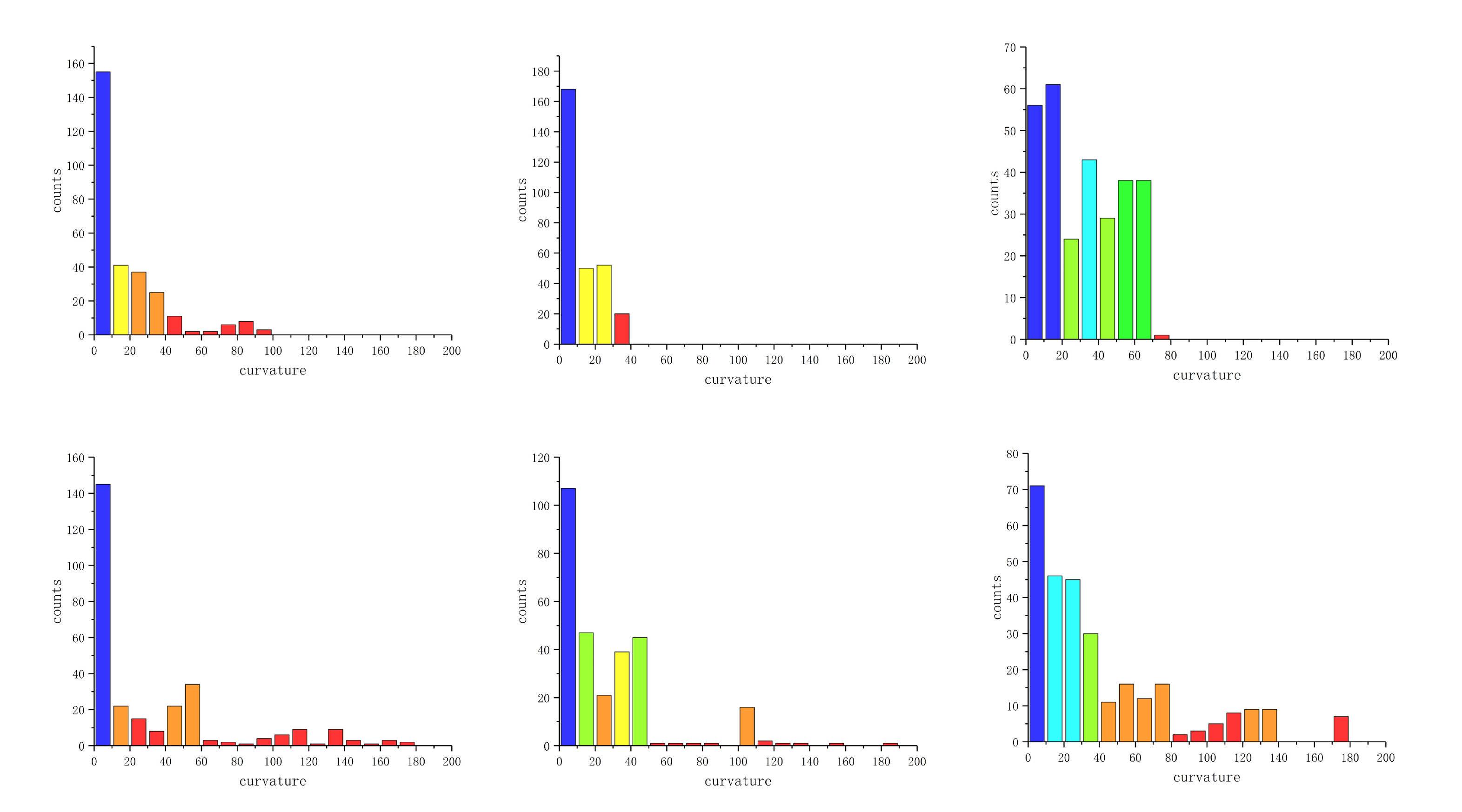}}
	\caption{WSCHs of distribution point clouds of six pathological ECG signals; the first row from left to right represents L. B. B. B., R. B. B. B., A. P.; the second row from left to right represents P. V. C., F. V. N., V. F.}
	\label{fig:curvatureothers}
\end{figure}

We calculate WSC of each point to precisely characterize the differences of neighborhood information of different point clouds on $SPD(3)$. WSCHs are also formed, as shown in Figure \ref{fig:curvatureothers}.

WSC sequences corresponding to seven kinds of ECG signals are almost located at $[0,200]$, as shown in Figure \ref{fig:histogram1} and Figure \ref{fig:curvatureothers}. By equation (\ref{cur inequality}), it can be inferred that the neighborhood information of most points in the Euclidean point clouds do not have more than two-dimensional degradation, which shows the effectiveness of the FFT embedding method and the selected parameters. 

In the histogram of normal ECG signal and A. P., their columns are evenly distributed horizontally and the columns of A. P. fluctuates more violently. The histograms of P. V. C., F. V. N. and V. F. are less evenly distributed horizontally and their fluctuation of columns are similar. The columns of L. B. B. B. and R. B. B. B. are concentrated in the smaller part of the WSC values. To identify these seven ECG signals more accurately, we calculate their WSCD. 

\begin{figure}[htbp]
	\centerline{\includegraphics[width=1\linewidth]{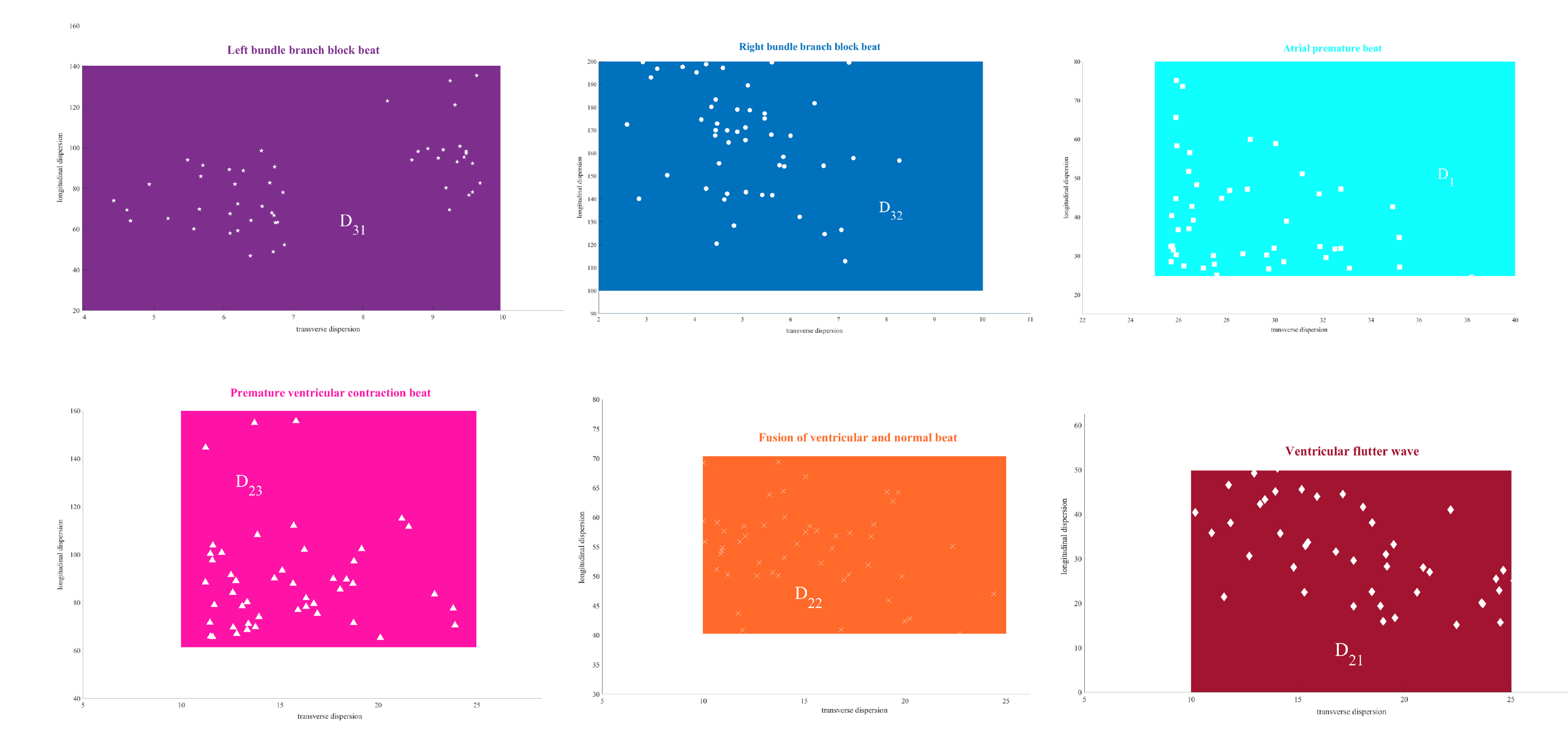}}
	\caption{WSCD of seven kinds of ECG signals; the first row from left to right represents L. B. B. B., R. B. B. B., A. P.; the second row from left to right represents P. V. C., F. V. N., V. F.}
	\label{fig:somepoints}
\end{figure}

Define Wasserstein scalar curvature dispersion plane:
\begin{equation*}
	\mathcal{D}=\left\{(cur_1(m,b,s),cur_2(m,b,s))\mid -\infty < cur_1(m,b,s)< +\infty, -\infty < cur_2(m,b,s)< +\infty \right\}.
\end{equation*}
Figure \ref{fig:somepoints} calculates WSCD of seven ECG signals. Transverse curvature dispersion and longitudinal curvature dispersion reveal the length and the fluctuation of QRS complex respectively. The transverse curvature dispersion gets larger with increasing width of QRS complex and the longitudinal curvature dispersion turns bigger with more violent fluctuation of QRS complex.
\begin{itemize}
	\item  QRS complex in normal ECG signal and A. P. are most regular and their transverse curvature dispersion are both largest. Notice that the lesion in the atria causes some changes in QRS complex, hence the longitudinal curvature dispersion of A. P. is higher than normal heartbeats. Let $D_0=(25,200]\times [0,25]$ denote normal status of heart and $D_1=(25,90]\times (25,+\infty)$ denote atrial abnormal.
	\item  QRS complex in V. F., F. V. N. and P. V. C. are not so wide and their fluctuation gradually expand, hence the transverse curvature dispersion of these three kinds of heartbeats are moderate and their longitudinal curvature dispersion becomes bigger and bigger. Because these three kinds of heartbeats are caused by ventricular lesions, their longitudinal dispersion have some similarities. Let $D_{21}=(10,25]\times [0,50]$ denote V. F., $D_{22}=(10,25]\times [40,70]$ denote V. F. N. and $D_{23}=(10,25]\times [60,+\infty)$ denote P. V. C.. Futhermore, let $D_2=(10,25]\times [0,+\infty)$ denote ventricular abnormal.
	\item  QRS complex in L. B. B. B. and R. B. B. B. are usually widest, hence their transverse curvature dispersion are smallest. In addition, the existence of two R peaks in R. B. B. B. is more obvious than in L. B. B. B., which results in more fluctuation in QRS and the larger longitudinal curvature dispersion. Let $D_{31}=[0,10)\times [0,140]$ denote L. B. B. B. and $D_{32}=[0,10)\times [100,\infty)$ denote R. B. B. B.. In addition, let $D_3=[0,10)\times [0,+\infty)$ denotes bundle branch block area.
\end{itemize}

For those heartbeats which land in $D_4=\mathcal{D}-\cup_{j=0}^{3} D_j$, we think these heartbeats are from other abnormal areas. Thus, we derive a symptom description domain partition $\mathcal{D}=\cup_{j=0}^{4} D_j$. Given a ECG signal $T_i$, the auxiliary diagnostic analysis of heart disease is as follows:
\begin{itemize}
	\item If $cur_{T_i}(m,b,s)\in D_0$, we think $T_i$ is normal.
	\item If $cur_{T_i}(m,b,s)\in D_1$, we think $T_i$ is A. P..
	\item If $cur_{T_i}(m,b,s)\in D_2 \cap D_{21}-D_{22}\cup D_{23}$, we think $T_i$ is V. F.. If $cur_{T_i}(m,b,s)\in D_2 \cap D_{22}-D_{21}\cup D_{23}$, we think $T_i$ is V. F. N.. If $cur_{T_i}(m,b,s)\in D_2 \cap D_{23}-D_{21}\cup D_{22}$, we think $T_i$ is P. V. C.. If $cur_{T_i}(m,b,s)\in D_{2j}\cap D_{2k},~1\leq j,k\leq 3$, we think $T_i$ has the pathological features of both $D_{2j}$ and $D_{2k}$. In this case, we cannot classify $T_i$, but we can label it ventricular abnormal. 
	\item If $cur_{T_i}(m,b,s)\in D_3 \cap D_{31}-D_{32}$, we think $T_i$ is L. B. B. B.. If $cur_{T_i}(m,b,s)\in D_3 \cap D_{32}-D_{31}$, we think $T_i$ is R. B. B. B.. If $cur_{T_i}(m,b,s)\in D_{31}\cap D_{32}$, we cannot classify $T_i$, but we can label it bundle branch block. 
	\item If $cur_{T_i}(m,b,s)\in D_4$, $T_i$ cannot be classified.
\end{itemize}	
Now we give our algorithm to show how to classify the ECG signals and carry out the auxiliary diagnosis. 

\subsection{WSCEC algorithm}
Let $\mathcal{T}=\{T_i\}_{i=1}^N=\cup_{j=0}^{r} \mathcal{T}_j=\cup_{j=0}^{4} \mathcal{Q}_j$ be the set of the given ECG signal where $\mathcal{T}_0$ is the set of normal ECG in $\mathcal{T}$, $\mathcal{T}_j$ is the set of ECG with $j$th pathological feature in $\mathcal{T}$ ($1\leq j \leq r$), $\mathcal{Q}_j$ is the original heartbeat set which is corresponding to symptom description domain $D_j$. Then WSCEC algorithm is shown in Algorithm \ref{alg:classification}.

For the output of algorithm \ref{alg:classification}, $\tilde{\mathcal{T}}_0$ represents the healthy heartbeat set, $\tilde{\mathcal{T}}_j$ represents the heartbeat set with $j$th pathological feature, $1\leq j \leq r$, and $\tilde{\mathcal{T}}_{r+1}$ represents the unclassified heartbeat set with label of lesion area. $\tilde{\mathcal{Q}}_j$ denotes the classified heartbeat set which is corresponding to symptom description domain $D_j$. Thus, for a unclassified heartbeat $T_i \in \tilde{\mathcal{T}}_{r+1}$, although we cannot know the exactly type of $T_i$, we can also know which part of the heart is abnormal.

\begin{algorithm}[H]
	\label{alg:classification}
	\caption{WSCEC algorithm} 
	\hspace*{0.02in} {\bf Input:} 
	ECG set $\mathcal{T}$; parameter $k,~m,~s,~\epsilon$\\
	\hspace*{0.02in} {\bf Output:} Classification result $\mathcal{T}=\cup_{j=0}^{r+1} \tilde{\mathcal{T}}_j\cup\tilde{\mathcal{Q}}_4=\cup_{j=0}^{4} \tilde{\mathcal{Q}}_j$
	\begin{algorithmic}[1]
		\State Choose standard normal ECG signal $T_{s}$
		\State For every ECG signal $T_{i}$ in $\mathcal{T}$, acquire point cloud $S_{F}(T_i)$ after FFT embedding by equation (\ref{fftpoint})
		\State Acquire point cloud $S_{SPD}(T_i,k)$ of $T_{i}$ and $S_{SPD}(T_{s},k)$ of $T_{s}$ by $kNN$ algorithm and equation (\ref{local mean})
		\State Calcuate scalar curvature at each point in $S_{SPD}(T_{s},k)$ by equation (\ref{Curve scalar}), take $b$ as the minimum of the max scalar curvature and $\frac{3d(d-1)}{\epsilon}$
		\State Calcuate scalar curvature at each point in $S_{SPD}(T_i,k)$ by equation (\ref{Curve scalar}) and obtain curvature histogram $H(m,b)$ by equation ($\ref{histogram}$)
		\State Calculate Wasserstein scalar curvature dispersion $cur_{T_i}(m,b,s)=(cur_1(m,b,s),cur_2(m,b,s))$ of $T_{i}$ by equation (\ref{cureq})
		\State Give the classification result $\mathcal{T}=\cup_{j=0}^{r+1} \tilde{\mathcal{T}}_j\cup\tilde{\mathcal{Q}}_4=\cup_{j=0}^{4} \tilde{\mathcal{Q}}_j$ by symptom description domain partition $\mathcal{D}=\cup_{j=0}^{4} D_j$
		\State \Return $\mathcal{T}=\cup_{j=0}^{r+1} \tilde{\mathcal{T}}_j\cup\tilde{\mathcal{Q}}_4=\cup_{j=0}^{4} \tilde{\mathcal{Q}}_j$
	\end{algorithmic}
\end{algorithm}

\section{Digital Experiment}\label{section 4}

In this section, the main numerical results are introduced. The sampled data comes from MIT-BIH Arrhythmia Database \cite{Marsanova L}. We sample $5000$ heartbeats with distinguishing features, including $2500$ normal heartbeats (N), $1000$ premature ventricular contraction heartbeats (P. V. C.), $450$ left bundle branch block heartbeats (L. B. B. B.), $450$ right bundle branch block heartbeats (R. B. B. B.), $200$ atrial premature heartbeats (A. P.), $200$ fusion of ventricular and normal heartbeats (F. V. N.) and $200$ ventricular flutter heartbeats (V. F.).

\begin{figure}[htbp]
	\centerline{\includegraphics[width=1\linewidth]{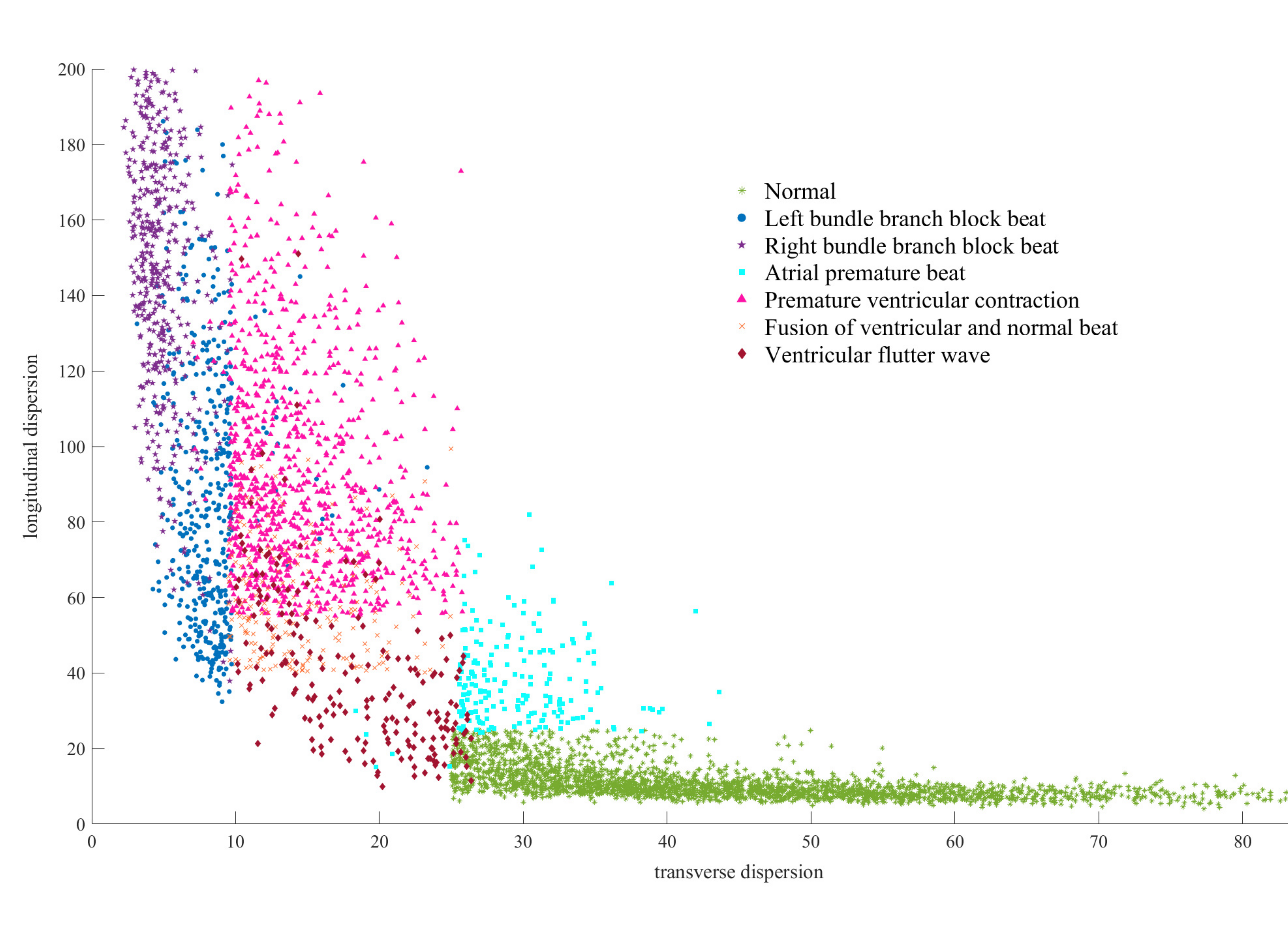}}
	\caption{Wasserstein scalar curvature dispersion of seven kinds of ECG signals}
	\label{fig:dispersion}
\end{figure}

Figure \ref{fig:dispersion} calculates WSCD of the $5,000$ segment ECG signals sampled, the parameters are $\epsilon=0.09,~b=\frac{3d(d-1)}{\epsilon}=200,~m=1$, and $s=0$. Note that there are overlaps between bundle branch block heartbeats and abnormal ventricular heartbeats such as F. V. N. and P. V. C., this may because all these heartbeats have widened QRS complex and some of them are indeed pretty similar. We introduce true positive rate (TPR) and noise removal rate (NRR) to show the efficiency of our symptom description domain partition.

Let $\mathcal{T}=\cup_{j=0}^{4} \mathcal{Q}_j$ be the original ECG set and $\mathcal{T}=\cup_{j=0}^{4} \tilde{\mathcal{Q}}_j$ be the set after classification. Then for all $0\leq j \leq 4$,
\begin{equation}
	\begin{aligned}
		{\rm TPR}_{j} & = \frac{|\tilde{\mathcal{Q}}_j\cap \mathcal{Q}_j|}{|\mathcal{Q}_j|}, \\
		{\rm NRR}_{j} & = 1-\frac{|\tilde{\mathcal{Q}}_j\cap \cup_{k\neq j} \mathcal{Q}_k|}{|\cup_{k\neq j} \mathcal{Q}_k|},
	\end{aligned}
\end{equation}
where $|\cdot|$ denotes the cardinality or size of a finite set.

${\rm TPR}_{j}$ describes the accuracy of original ECG signals with lesion area $j$ preserved by the new classification. ${\rm NRR}_{j}$ represents the success rate of removing ECG signals except $j$. It can be intuitively understood that higher TPR and NRR represent better classification ability. Table \ref{tab:result} shows the classification accuracy of symptom description domain partition. Intuitively, the normal ECG can be very accurately separated from other heartbeats with pathological features.

\begin{table}[!htp]
	\centering
	\setlength\tabcolsep{0.04pt}
	\setlength{\abovecaptionskip}{0cm} 
	\setlength{\belowcaptionskip}{0.3cm}
	\renewcommand{\arraystretch}{1.2}
	\caption{Classification result of symptom description domain partition}
	\setlength{\tabcolsep}{0.5mm}{
		\begin{tabular}{|c|c|c|ccc|cc|c|}
			\hline
			\multirow{2}{*}{ECG type}                                                                 & Normal & Atrial abnormal & \multicolumn{3}{c|}{Ventricular   abnormal}                           & \multicolumn{2}{c|}{Bundle   branch block}     & Unclassified \\ \cline{2-9} 
			& N      & A. P.           & \multicolumn{1}{c|}{P. V. C.} & \multicolumn{1}{c|}{F. V. N.} & V. F. & \multicolumn{1}{c|}{L. B. B. B.} & R. B. B. B. & Null         \\ \hline
			{\begin{tabular}[c]{@{}c@{}}Original size\end{tabular}}               & 2500   & 200             & \multicolumn{3}{c|}{1400}                                             & \multicolumn{2}{c|}{900}                       & 0            \\ \hline
			{\begin{tabular}[c]{@{}c@{}}Classified size\end{tabular}} & 2513      & 212               & \multicolumn{3}{c|}{1305}                                                & \multicolumn{2}{c|}{970}                         & 0           \\ \cline{2-9} \hline
			TPR                                                                                                & 99.96\%      & 95.50\%               & \multicolumn{3}{c|}{91.36\%}                                                & \multicolumn{2}{c|}{97.78\%}                         & 0            \\ \hline
			NRR                                                                                                & 99.44\%     & 99.56\%               & \multicolumn{3}{c|}{99.35\%}                                                & \multicolumn{2}{c|}{97.80\%}                         & 0            \\ \hline
	\end{tabular}}
	\label{tab:result}
\end{table}

\begin{figure}[htbp]
	\centerline{\includegraphics[width=1.1\linewidth]{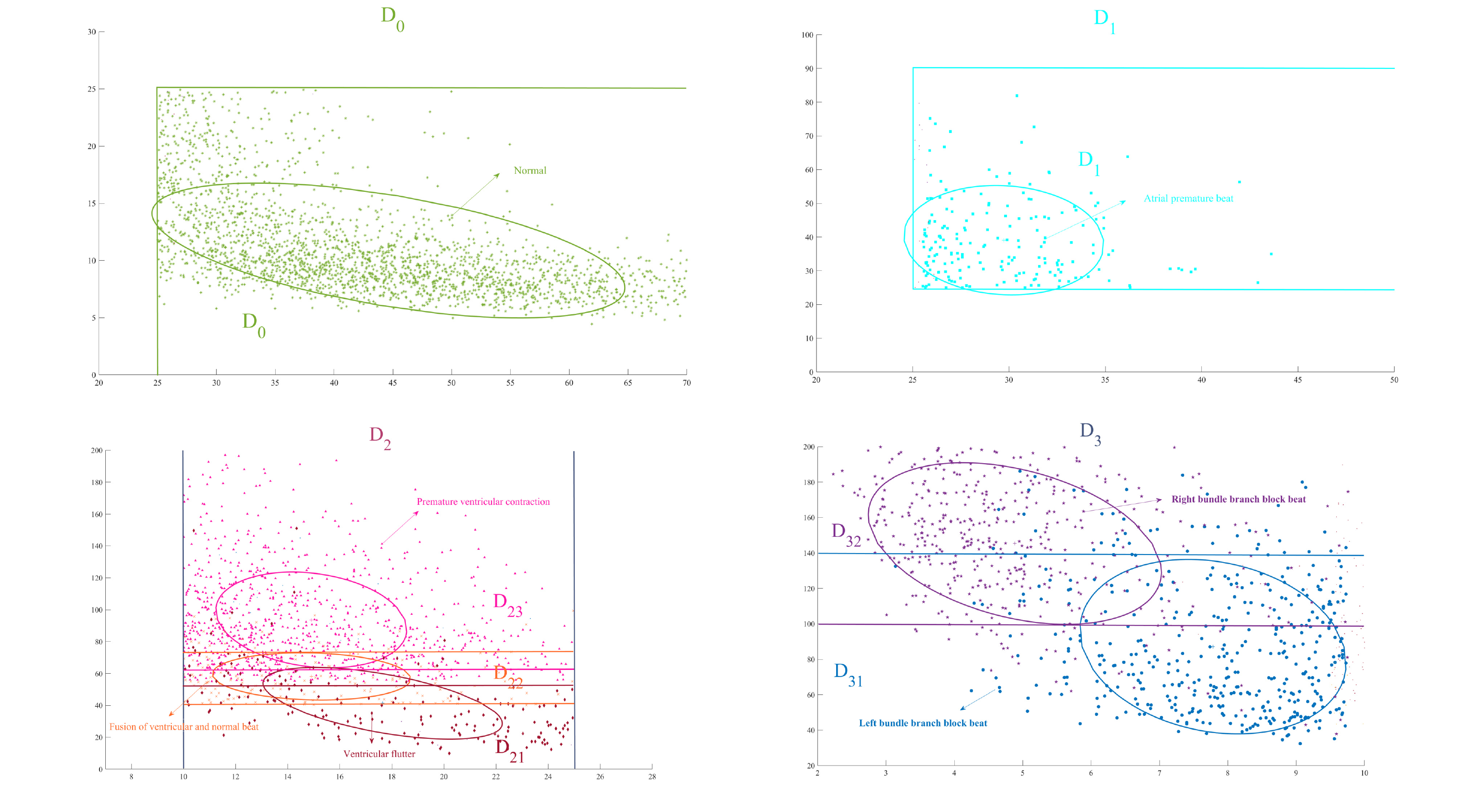}}
	\caption{Principal component analysis result}
	\label{fig:PCA}
\end{figure}

To show the classification results of each specific disease and the validity of symptom description domain partition, we use principal component analysis (PCA) to obtain the confidence elliptic region\cite{Stevens J P}, as shown in Figure \ref{fig:PCA}. Notice that almost every confidence elliptic region is contained in the corresponding symptom description domain, hence the selection of our symptom description domain partition has strong robustness.

Although the results from normal ECG signals are basically identical, ECG signals with different pathological features tend to have different degrees of change and ECG signals may vary even if they share same pathological feature. Thus, firstly, we infer the location of cardiac abnormalities such as atrium, ventricle and atrioventricular bundle. Secondly, we give the reference diagnostic results. For those ECG signals which we can infer the location of cardiac abnormality but cannot make sure the exactly disease, doctors needs to make further diagnosis. Furthemore, we can estimate the severity of pathology according to the longitudinal dispersion.

\section{Conclusions and Future Works}\label{section 5}

In this paper, a novel ECG assisted classification algorithm based on Wasserstein scalar curvature is proposed. By introducing Wasserstein scalar curvature, WSCEC algorithm more accurately describes the neighborhood information differences of point clouds obtained after FFT embedding, and realizes the classification of seven kinds of single-lead ECG with different pathological features. The accuracy and interpretability of WSCEC algorithm are strong.

WSCEC algorithm is an original attempt to incorporate geometric invariants into medical research, and it is expected to be further applied to other big databases and the standard $12$-lead ECG auxiliary diagnosis. Meanwhile, WSCEC algorithm can also be applied to a variety of signal research such as signal identification and Electroencephalogram (EEG) analysis. How to choose a more reasonable signal embedding way and how to further combine with topological data analysis to comprehensively investigate the signal from the whole and local to get more efficient algorithms need further research.

\end{document}